\documentclass[aps,prd,apj,10pt,twocolumn,amsfonts,amssymb,amsmath,showpacs,letterpaper,superscriptaddress,nofootinbib,altaffilletter,fontenc,orcidlink]{revtex4-2}
\usepackage{lineno}
\usepackage[utf8]{inputenc}
\usepackage[T1]{fontenc}
\usepackage{graphicx}	
\graphicspath{ {./Figures/} }
\usepackage{caption}
\usepackage{subcaption}
\usepackage{amsmath}	
\usepackage{amssymb}	
\usepackage{tabularx}
\usepackage{enumerate}
\usepackage{enumitem}
\usepackage{comment}
\usepackage{multirow}
\usepackage{float}
\usepackage[citecolor=blue,linkcolor=blue,colorlinks=true,urlcolor=blue]{hyperref}
\usepackage[misc]{ifsym}
\usepackage{orcidlink}
\usepackage{xcolor}

\usepackage{natbib}
\newcommand{\orcid}[1]{%
  \href{https://orcid.org/#1}{\includegraphics[height=1.8ex]{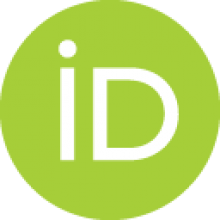}}%
}
\begin{document}
\title{Estimating the Lensing Probability for Binary Black Hole Mergers in AGN disk by Using Mismatch Threshold}

\author{Wen-Long Xu,\orcid{0009-0003-9792-9325}}

\affiliation{Department of Astronomy, School of Physics, Huazhong University of Science and Technology, Luoyu Road 1037, Wuhan, 430074, China}

\author{Yu-Zhe Li,\orcidlink{0009-0002-9215-5618}
}
\affiliation{Department of Astronomy, School of Physics, Huazhong University of Science and Technology, Luoyu Road 1037, Wuhan, 430074, China}

\author{Yi-Gu Chen,\orcidlink{0000-0002-8043-6650}}
\affiliation{Department of Astronomy, School of Physics, Huazhong University of Science and Technology, Luoyu Road 1037, Wuhan, 430074, China}

\author{Hui Li,\orcidlink{0000-0003-3556-6568}
}
\email{hli@lanl.gov}
\affiliation{Theoretical Division, Los Alamos National Laboratory, Los Alamos, NM 87545, USA}

\author{Wei-Hua Lei,\orcidlink{0000-0003-3440-1526}
}
\email{leiwh@hust.edu.cn}
\affiliation{Department of Astronomy, School of Physics, Huazhong University of Science and Technology, Luoyu Road 1037, Wuhan, 430074, China}

\begin{abstract}
Stellar-mass binary black holes (BBH) may form, evolve, and merge within the dense environments of active galactic nuclei (AGN) disks, thereby contributing to the BBH population detected by gravitational wave (GW) observatories. 
Mergers occurring in AGN disks may be gravitationally lensed by the central supermassive black hole. 
Such lensing presents a potential probe of BBH formation channels through its probability. 
In this work, we calculate the AGN lensing probability of LIGO sources embedded in AGN disk by relating mismatch threshold to the signal-to-noise ratio. 
The AGN lensing probability of GW190521-like event will be $\sim 3\%$ for O3 PSD, $\sim 6\%$ for Aplus PSD, and $\sim 33\%$ for ET PSD. 
If AGNs are indeed the primary formation channel for BBHs, we could quantify the probability of detecting the lensed GW events in such a scenario. 
The non-detections, on the other hand, will place stricter constraints on the fraction of AGN disk BBHs and even the birthplaces of BBH mergers.
\end{abstract}

\date[\relax]{Dated: \today }

\maketitle

\section{Introduction}
\label{sec:intro}

In 2015, LIGO detected the first binary black hole (BBH) merger event, known as GW150914 \citep{2016PhRvL.116m1103A}. 
Two years later, another significant event was observed – the merger of a binary neutron star, GW170817 \citep{2017PhRvL.119p1101A}. This event was accompanied by the electromagnetic counterpart of GRB 170817A \citep{2017ApJ...848L..12A,2017ApJ...848L..14G} and kilonova AT 2017gfo/DLT17ck \citep{2017ApJ...848L..24V}. Gravitational waves from two neutron star–black hole (BH) coalescences were also observed in January 2020 \citep{2021ApJ...915L...5A}. 
The network of ground-based gravitational-wave (GW) detectors, after four observing runs, has observed more than 300 events, most of which are merging BBHs \citep{2019PhRvX...9c1040A,2021PhRvX..11b1053A,2023PhRvX..13d1039A,GWTC-2.1:catalog,2025arXiv250818080T,LIGO2025}. 
More GW events are expected to be detected in future observation runs and with new detectors such as LIGO-india \citep{2022CQGra..39b5004S}, Deci-hertz Interferometer GW Observatory (DECIGO, \citep{2021PTEP.2021eA105K}), Einstein Telescope (ET, \citep{2010CQGra..27s4002P}), and Cosmic Explorer (CE, \citep{2023arXiv230613745E}). 
This rapidly growing population motivates detailed studies of BBH formation channels.

The proposed scenarios include chemically homogeneous evolution of isolated binary stars \citep{2016MNRAS.460.3545D}, common-envelope ejection \citep{1976IAUS...73...75P,2013A&ARv..21...59I}, envelope expansion \citep{2018PhRvL.120z1101T}, chemically homogeneous evolution in a tidally distorted binary \citep{2016MNRAS.460.3545D}, evolution of triple or quadruple systems \citep{2017ApJ...841...77A,2017ApJ...846L..11L}, gravitational capture \citep{2009MNRAS.395.2127O}, dynamical evolution in open clusters \citep{2017MNRAS.467..524B} and dense star clusters \citep{2000ApJ...528L..17P}, dynamical interaction in gas-rich nuclear regions \citep{2012MNRAS.425..460M}, and the evolution of Population III binary stars \citep{2002ApJ...571...30S}, among others. 
However, the observation of BBH merger event GW231123 indicates that BBH masses are consistent with intermediate mass, and both black hole spins are consistent with very high spin parameters \citep{2025ApJ...993L..25A}. 
These two distinctive properties pose challenges to conventional formation channels.

Recent studies suggested that GW231123 may originate from active galactic nuclei (AGN channel). 
For example, Ref. \citep{2025arXiv250808558B} proposed that GW231123 could be formed via accretion onto Population III remnants in an AGN disk, whereas Ref. \citep{2025arXiv250813412D} suggested that it may arise from hierarchical mergers within AGN disks. 
Although electromagnetic counterpart observations could, in principle, provide constraints on the localization and environment of GW231123 \citep{2017ApJ...835..165B,2024ApJ...961..206C}, whether such counterparts are detectable remains uncertain \citep{2019ApJ...884L..50M,2021ApJ...916..111K,Yuan2025}. 
Thus, a single observational signature may be insufficient to uniquely identify the formation environment of BBHs, highlighting the need for multiple diagnostic approaches. 
In this context, we propose that AGN lensing may serve as a complementary diagnostic tool.

The basis for this proposal is that the complex environment of AGN may lead to diverse AGN lensing signatures, which in turn underpin their diagnostic potential. 
For example, BBHs prior to merger, while orbiting the central supermassive black hole (SMBH), may undergo repeated gravitational lensing events \citep{2020PhRvD.101h3031D,2021PhRvD.104j3011Y}, leading to periodic spikes in GW amplitude. 
As the BBH passes near caustics, its waveforms could be distorted due to diffraction \citep{2025PhRvD.112d3544E}. 
The presence of a lensing object may also cause the GW to experience the gravitational spin Hall effect during propagation \citep{2018PhRvD..98f1701Y,2021PhRvD.103d4053A,2024PhRvD.109l4045O,2024MNRAS.535L...1O}, thereby altering its polarization \citep{2019PhRvD.100f4028H}. 
Moreover, the stellar field \citep{2018PhRvD..98j3022C,2023SCPMA..6639511S} surrounding the BBH and galaxy substructure \citep{2023MNRAS.525.4149L,2025arXiv250821262S} may also play a role in shaping the AGN lensing signature.

Although the AGN lensing signatures are diverse, they are not beyond investigation. 
Previous work has established a framework for lensed GW signal search and identification \citep{2021ApJ...923...14A,2024ApJ...970..191A,2025arXiv251216347T}, which provides a foundation for such investigations. 
Generally, each GW event is first identified by matched-filtering search pipelines \citep{2019PhRvX...9c1040A,2021PhRvX..11b1053A,2023PhRvX..13d1039A,GWTC-2.1:catalog,2025arXiv250818080T,LIGO2025}. 
The assessment of potential lensing is performed subsequently \citep{2019ApJ...874L...2H,2020arXiv200712709D,2021ApJ...923...14A,2023MNRAS.526.3832J,2024ApJ...970..191A,2025arXiv251216347T,2025arXiv251217631G,2025arXiv251219077C,2025arXiv251217550H}, through the identification of lensing signatures \citep{2018arXiv180707062H,2021PhRvD.103f4047E,2023PhRvD.107l3014L} using model selection methods \citep{2021MNRAS.506.5430J,2021ApJ...908...97L,2025arXiv251015463H,2025arXiv250720256W}. 
Reference \citep{2024MNRAS.531..764M} employed a comprehensive approach combining Bayesian inference and matched-filtering analysis to systematically investigate lensing effects. 
Their results indicate that the maximum match across the $M_{\rm L}$ – $y$ plane effectively captures the similarity between lensed and unlensed waveforms. 
This motivates us to employ the match-filtering method to investigate and quantify the distribution of waveform deviations induced by AGN lensing across the AGN disk. 

Existing research indicates that waveform deviations can be quantified using mismatch \citep{2008PhRvD..78l4020L,2025PhRvD.112f4011T}. 
With a reasonable construction of the AGN lensing geometry, it is possible to map the distribution of waveform deviations induced by AGN lensing across the disk via the mismatch. 
Unfortunately, even if such waveform deviations are quantified, they may not necessarily be measurable. 
Recent studies show that the mismatch can be related to an indistinguishability signal-to-noise ratio (SNR) \citep{2025PhRvD.112f4011T}. 
Specifically, if the mismatch $\epsilon$ falls below a certain threshold (mismatch threshold hereafter), the intrinsic deviation between waveforms becomes smaller than the statistical uncertainty \citep{2025PhRvD.112f4011T}. 
In such cases, the waveforms are considered indistinguishable \citep{1992PhRvD..46.5236F,1993PhRvD..47.2198F,1994PhRvD..49.2658C,1998PhRvD..57.4566F,2008PhRvD..78l4020L,2009PhRvD..80d2005L,2010PhRvD..82b4014M}. 
This indistinguishability suggests two implications. 
First, AGN lensing effects need to be sufficiently pronounced to have a chance of being distinguished from other AGN-related effects. 
Second, for a merger event at a given disk location, if the AGN lensing-induced waveform deviation is indistinguishable, that event would not contribute to the overall AGN lensing probability. 
This offers a potential method for estimating the AGN lensing probability across the disk using the mismatch threshold and SNR. 

GW lensing rates have been derived from optical depth, adopting the Einstein radius serving as a key criterion \citep{2018MNRAS.474.2975D}. 
Using this approach, Ref. \citep{2025ApJ...979L..27L} found that, with $\sim 100$ observations, the lensing probability of observing at least one GW190521-like event is about $14\%$. 
However, the Einstein-radius criterion is not designed to reflect the specific distribution of BBHs within AGN disks, nor does it directly incorporate the effects of intrinsic binary parameters  \citep{2022PhRvD.105f3006L,2019ApJ...876..122Y}. 
Therefore, this motivates us to investigate the AGN lensing probability based on the mismatch threshold. 
This criterion, in turn, is shaped by several key factors: the source distance, the combination of binary mass, and the detector power spectral density (PSD), which constitutes the main subject of this study. 

This work is organized as follows. 
In Section \ref{sec:theory}, we present the geometric configuration of our AGN lensing model and match-filtering method. 
In Section \ref{sec:Calculation}, we present calculations of the waveform mismatch induced by AGN lensing and its disk distribution, together with the estimation of AGN lensing probability. 
Conclusions and discussions are given in Section \ref{sec:Conclusion&diskussion}. 

\section{Gravitational Lensing of BBH mergers in AGN disk and mismatches}
\label{sec:theory}
\begin{figure*}
    \includegraphics[width=0.95\linewidth]{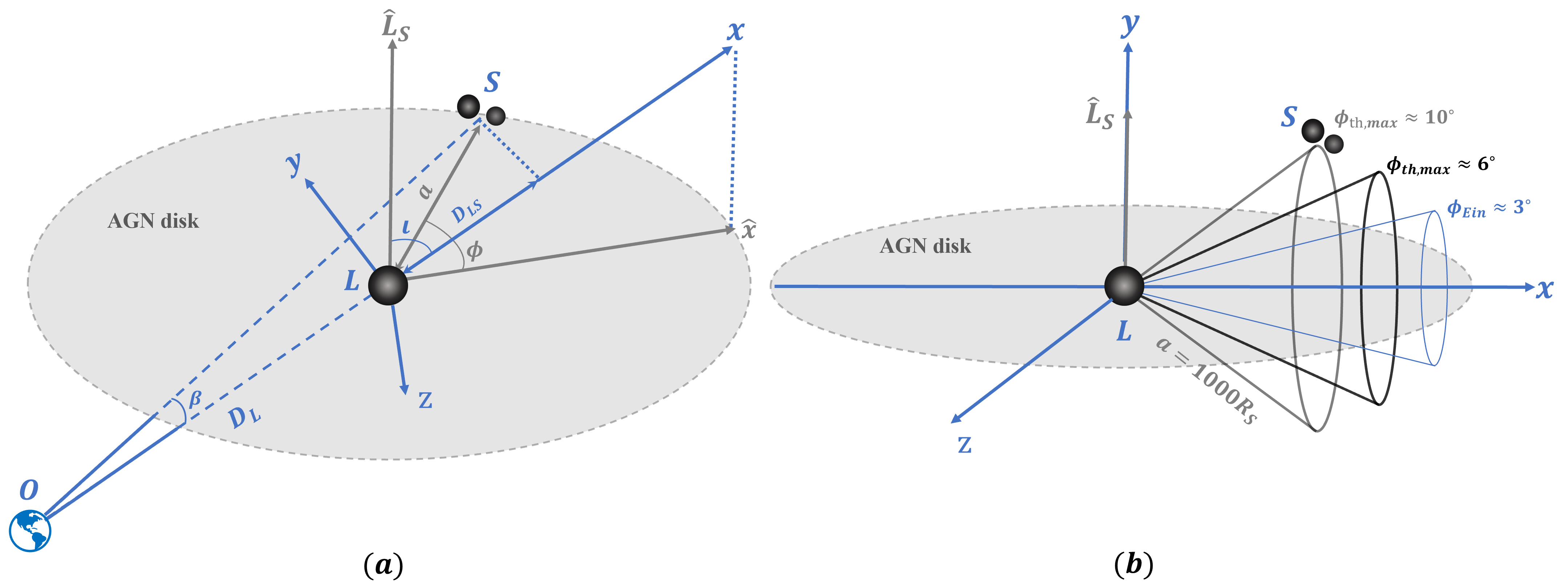}
    \caption{Schematic diagram of the lensing system. FIG. \ref{fig:lens_sys} $(a)$ showing a BBH (Source $S$) orbiting an SMBH (Lens object $L$). The gray-shaded region indicating the orbital plane of the source. When the BBHs are located in the conical region on the right, they will form a lensing system with the central SMBH and the observer $O$ on the left. $D_{\rm L}$, $D_{\rm S}$ and $D_{\rm LS}$ represent the distance from the observer to lens, the observer to source and the lens to source, respectively. $\beta$ is the angular position of the source. The position of the source in the disk is represented by $\phi$, $\iota$, and $a$, where $\phi$ denotes the orbital phase, $\iota$ indicates the angle between the source's orbital angular momentum ($\hat{L}_{\rm S}$) and optical axis ($x$-axis), while $a$ represents the semi-major axis of the source's orbit. Several cones with opening angles of $\phi_{\rm th,max}$ are also plotted to depict the threshold orbital phase. We get FIG. \ref{fig:lens_sys} $(b)$ by setting $\iota = 90^{\circ}$. $\phi_{\rm th,max} \approx 6^{\circ}$ and $10^{\circ}$ corresponds to different semi-major axis. The blue solid angle of $\phi_{\rm Ein} \approx 3^{\circ}$ is given by Einstein criterion.}
    \label{fig:lens_sys}
\end{figure*}

\begin{figure}
    \centering
    \includegraphics[width=\columnwidth]{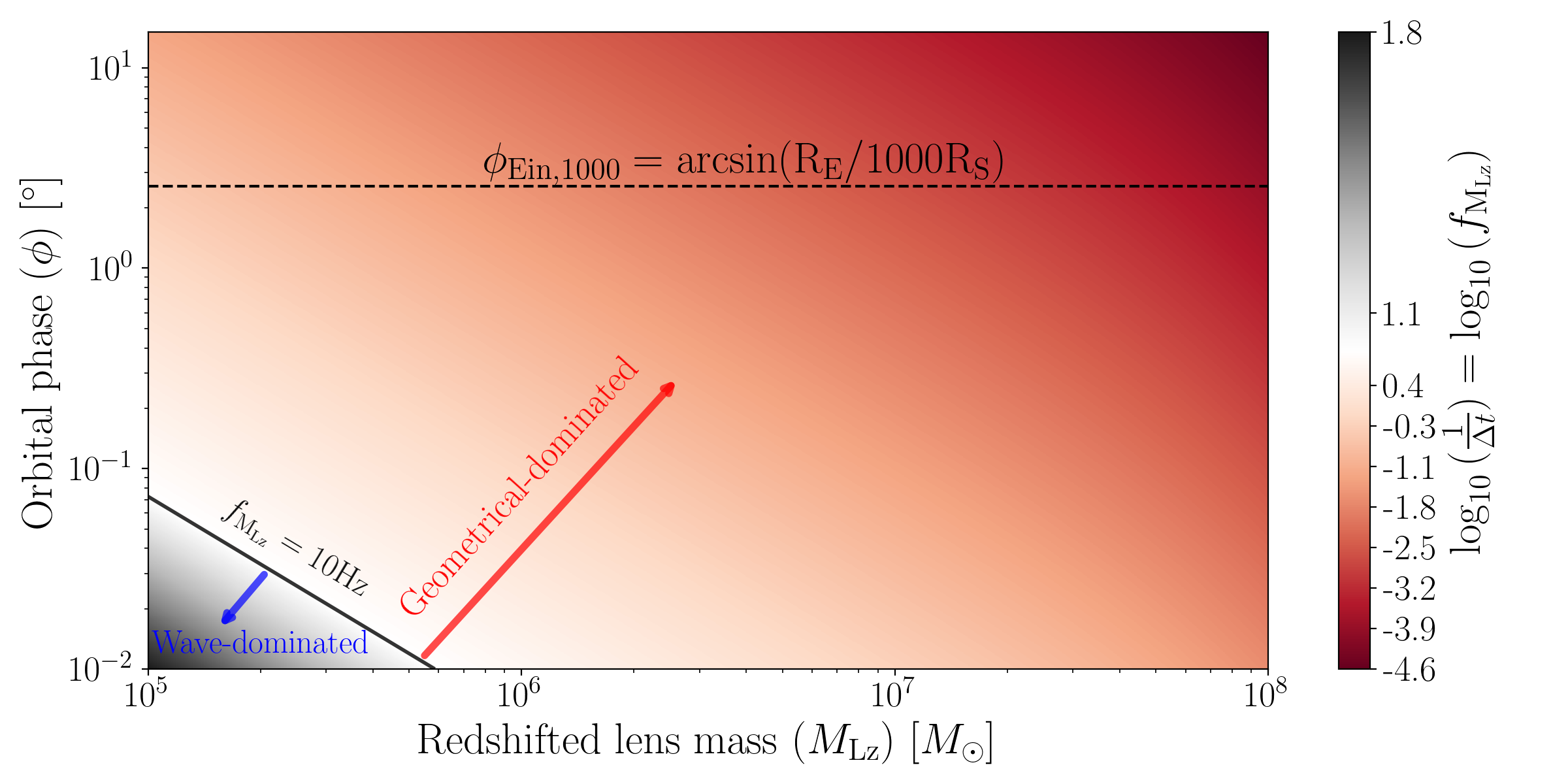}
    \caption{Contour plot of the characteristic frequency, $f_{\rm{M}_{\rm{Lz}}}$. Wave-optics dominated regime: bottom left region, where $f_{\rm{GW}} \sim f_{\rm{M}_{\rm{Lz}}}$. The frequency $f_{\rm{M}_{\rm{Lz}}}= 10$ $\rm{Hz}$ corresponds to the lower limit of the LIGO-Virgo sensitivity band. Geometrical-optics regime: upper right region, where $f_{\rm{GW}} \gg f_{\rm{M}_{\rm{Lz}}}$. For reference, we have plotted the Einstein criterion determined phase angles $\phi_{\rm{Ein,1000}}$ in dashed black line, when $\iota = 90^{\circ}$.  As seen from the dashed black line in the figure, the Einstein criterion related orbital phase fall entirely within the geometric optics limit.} 
    \label{fig:lensing_regime}
\end{figure}

The geometric configuration of the AGN lensing effect of the GW adopted in this work is depicted in FIG. \ref{fig:lens_sys} ($a$). 
Here, the source is the BBH, which is orbiting the SMBH that is considered as the lensing object $L$.
The reference frame $(x,y,z)$ considered here is centered on the lensing object $L$. 
The $x$-$\rm{axis}$ is defined by the line of sight between the observer and the lensing object $L$. 
The $x$-$y$ plane is defined by the plane containing both the $x$-$\rm{axis}$ and the source's orbital angular momentum vector $\hat{L}_{\rm S}$ around the object $L$ \citep{2019ApJ...887..210F}, the angle between the $y$-$\rm{axis}$ and $\hat{L}_{\rm S}$ is $90^{\circ} - \iota$. 
The source location can thus be written as
\begin{equation}
\left\{
             \begin{array}{lr}
             x(a,\phi,\iota)=a \cos \phi \sin \iota , &  \\
             y(a,\phi,\iota)=a \cos \phi \cos \iota , & \\
             z(a,\phi)=-a \sin \phi, &  
             \end{array}
\right.
\label{eq:frame}
\end{equation}
where $a$ is the semi-major axis of the source, $\phi$ is the orbital phase of the source and $\iota$ is the inclination angle between the line of sight and the orbital angular momentum of the source $\hat{L}_{\rm S}$, with $cos(\iota) = \boldsymbol{x} \cdot \boldsymbol{\hat{L}_{\rm S}}$, respectively. 
From FIG. \ref{fig:lens_sys} ($a$), Eq. \eqref{eq:frame} and \eqref{eq:beta}, we can see that the angular source position $\beta$ is determined by $a,\phi,\iota$. 
Specifically, when $\iota=0^{\circ}$, this corresponds to a face-on disk. The angular position of the source is $\beta = a/D_{\rm L}$, independent of the source’s orbital phase $\phi$, and the AGN lensing effect may disappears in such a configuration. 
On the other hand, for a perfectly aligned AGN lensing system, we will have $\iota = 90^{\circ}$, $\phi = 0^{\circ}$ and $\beta = 0^{\circ}$. 
The dimensionless angular source location of point mass lens model \citep{schneider1992gravitational} can be expressed as
\begin{eqnarray}
\eta(a,\phi,\iota)=\frac{\beta(a,\phi,\iota)}{\theta_{\rm Ein}(a,\phi,\iota)},
\label{eq:eta}
\end{eqnarray}
where
\begin{eqnarray}
\beta(a,\phi,\iota) = \frac{\sqrt{y^2(a,\phi,\iota)+z^2(a,\phi,\iota)}}{D_{\rm L}}
\label{eq:beta}
\end{eqnarray}
is the angular position of the source, and
\begin{eqnarray}
\theta_{\rm Ein}(a,\phi,\iota)=\sqrt{2R_{\rm S} \frac{D_{\rm LS}(a,\phi,\iota)}{D_{\rm L} D_{\rm S}}}
\label{eq:einstein_angle}
\end{eqnarray}
is the angular Einstein radius with $D_{\rm LS}=x(a,\phi,\iota)$ and $R_{\rm S} = 2GM_{\rm Lz}/c^2$ is the Schwarzschild radius of the lens with mass $M_{\rm Lz}$. 
The time delay between these two images is expressed as 
\begin{align}
    \Delta t(a,\phi,\iota,M_{\rm Lz}) =
    &\frac{4GM_{\rm Lz}}{c^3} [\frac{\eta(a,\phi,\iota) \sqrt{\eta^2(a,\phi,\iota) +4}}{2} + \nonumber \\    &\ln{(\frac{\sqrt{\eta^2(a,\phi,\iota)+4}+\eta(a,\phi,\iota)}{\sqrt{\eta^2(a,\phi,\iota)+4}-\eta(a,\phi,\iota)})}]. \label{eq:time_delay}
\end{align} 
Their magnifications are
\begin{eqnarray}
\mu_{\pm}(a,\phi,\iota)=\frac{1}{2} \pm \frac{\eta^2(a,\phi,\iota)+2}{2 \eta(a,\phi,\iota) \sqrt{\eta^2(a,\phi,\iota)+4}}.
\label{eq:magnification_factor}
\end{eqnarray} 
Following Ref. \cite{2003ApJ...595.1039T}, the amplification factor of the wave optic solution that depends on the coordinate parameters has the form of

\begin{align}
    F(f,a,\phi,\iota,M_{\rm Lz}) =& \exp\left\{\frac{\pi w}{4} + i \frac{w}{2}
    \left[\ln\left(\frac{w}{2}\right)-2\phi_{\rm m}(\eta)\right]\right\}\nonumber \\
    &\times \Gamma(1-i\frac{w}{2})_1F_1(i\frac{w}{2}, 1; i\frac{w\eta^2}{2}),
    \label{eq:transmission_factor}
\end{align}
where $\Gamma$ is the complex gamma function, $_1F_1$ is confluent hypergeometric function of the first kind, and 
\begin{eqnarray}
w(f,M_{\rm Lz}) = \frac{8\pi G M_{\rm Lz} f}{c^3}
\label{eq:w}
\end{eqnarray}
is the dimensionless frequency. $\phi_{\rm m}(a,\phi,\iota)$ is defined as
\begin{align}
    \phi_{\rm m}(a,\phi,\iota) =
    &\frac{1}{2}[\eta_{\rm m}(a,\phi,\iota)-\eta(a,\phi,\iota)]^2 \nonumber\\ -   
    &\ln{[\eta(a,\phi,\iota)]} 
    \label{eq:phi_m},
\end{align}
where
\begin{eqnarray}
\eta_{\rm m}(a,\phi,\iota)=\frac{1}{2} \{\eta(a,\phi,\iota)+[\eta^2(a,\phi,\iota)+4]^{1/2}\}.
\label{eq:eta_m}
\end{eqnarray}
In the geometrical optics limit ($w\gg 1$), we have
\begin{align}
    F(f,a,\phi,\iota,M_{\rm Lz}) =& |\mu_{+}(a,\phi,\iota)|^{1/2} - \nonumber\\
    &i |\mu_{-}(a,\phi,\iota)|^{1/2}e^{2\pi i f\Delta t(a,\phi,\iota,M_{\rm Lz})}.
    \label{eq:F_GO}
\end{align}

Following Ref. \cite{2024MNRAS.531..764M}, we use the relation $f_{M_{\rm Lz}} \equiv 1/\Delta t$ to determine which region of the parameter space, the geometrical optical regime, the wave-dominated regime, or the long-wavelength regime, our sources reside in. 
Following Ref. \citep{2023PhRvD.108h4033B}, we make a smooth connection from wave optical solution (see Eq. \eqref{eq:transmission_factor}) to geometrical optic limits (i.e. Eq. \eqref{eq:F_GO}). 

For a given unlensed waveform template \textbf{$\tilde{h}_{\rm UL}(f)$}, the lensed waveform \textbf{$\tilde{h}_{\rm L}(f)$} is defined as the product of the  amplification factor and the unlensed waveform, as
\begin{align}
    \tilde{h}_{\rm L}(f,a,\phi,\iota,M_{\rm Lz})=F(f,a,\phi,\iota,M_{\rm Lz})\tilde{h}_{\rm UL}(f).
    \label{eq:waveform}
\end{align}
The match ($\mathcal{M}$) between these two waveforms provides a method to quantify their similarity, which has a value of zero to unity \citep{1994PhRvD..49.2658C}.  
The match between lensed and unlensed waveforms are 
\begin{align}
    \mathcal{M}(h_{\rm UL},h_{\rm L})=\max _{\phi_{0}, t_{0}} \frac{\langle h_{\rm L}|h_{\rm UL}\rangle}{\sqrt{\langle h_{\rm L}|h_{\rm L}\rangle \langle h_{\rm UL}|h_{\rm UL}\rangle}},
    \label{eq:match}
\end{align}
where 
\begin{equation}
    \langle h_{\rm L}|h_{\rm UL}\rangle \equiv 2\int
    \frac{\tilde{h}^\ast_{\rm L}(f)\tilde{h}_{\rm UL}(f) + \tilde{h}_{\rm L}(f)\tilde{h}^\ast_{\rm UL}(f)}{S_{\rm n}(f)} df
    \label{eq:inner_product}
    \vspace{1em}
\end{equation}
is the noise weighted inner product between lensed and unlensed waveforms, with $S_{\rm n}(f)$ as the one-sided noise power spectral density. 
As a result, the AGN lensing induced waveform deviation is represented by the mismatch between the lensed and unlensed waveforms 
\begin{equation}
    \epsilon(f,a,\phi,\iota,M_{\rm Lz}) = 1 - \mathcal{M}. 
    \label{eq:mismatch}
\end{equation}
The mismatch threshold $\epsilon_{th}$ is used to determine whether the two waveforms can be distinguished \citep{1992PhRvD..46.5236F,1993PhRvD..47.2198F,1994PhRvD..49.2658C,1998PhRvD..57.4566F,2008PhRvD..78l4020L,2009PhRvD..80d2005L,2010PhRvD..82b4014M},
\begin{equation}
    \epsilon_{\rm th}(f,a,\iota,M_{\rm Lz};\phi_{\rm th}) = \frac{1}{\langle h_{\rm L}|h_{\rm L}\rangle + \langle h_{\rm UL}|h_{\rm UL}\rangle}
    \simeq \frac{1}{2\rho^2}.
    \label{eq:threshold}
\end{equation}
where parameter $\phi_{\rm{th}}$ represents the threshold orbital phase corresponding to the mismatch threshold $\epsilon_{\rm th}$, and $\rho$ refers to the SNR. 
The square of the SNR of the lensed waveform is defined as 

\begin{equation}
    \rho^2 = 4\int{\frac{|\tilde{h}_{\rm{L}}(f)|^2}{S_{\rm{n}}(f)}df}.
    \label{eq:signal_to_noise_ratio}
\end{equation}


\section{Mismatch and Lensing Probability} 
\label{sec:Calculation}
In this section, given the dependence between the mismatch threshold and the SNR, we first calculate the SNR, then compute the mismatch arising from the AGN lensing effect, and finally estimate the lensing probability using the mismatch threshold $\epsilon_{\rm th}$.

\subsection{Signal-to-Noise Ratio}
\label{subsec:Signal-to-Noise Ratio}
In this subsection, we employ PyCBC \citep{2005PhRvD..71f2001A,2012PhRvD..85l2006A,2014PhRvD..90h2004D,2016CQGra..33u5004U,2017ApJ...849..118N,2024zndo..10473621N} to calculate the SNR. 
Our calculations primarily consider the following dependence: source distance, different mass combinations, and three different detector sensitivities. 

The mass of sources observed by LIGO range  from several solar masses to few hundred solar masses, with the merger rate following a broken power-law distribution \citep{2020PhRvL.125j1102A}. 
Accordingly, we select $20 M_{\odot} +20M_{\odot}$ and $35 M_{\odot} +30M_{\odot}$ as representative examples for our calculations. 
Furthermore, considering that higher mass events may result from hierarchical mergers \citep{2020ApJ...902L..26F}, we also adopt the mass combination corresponding to the GW190521-like event (e.g $85 M_{\odot} +66M_{\odot}$) \citep{2020PhRvL.125j1102A}. 
Regarding the PSDs, our focus lies on how improvements in PSD affect the SNR, as well as the potential enhancement in SNR from future multi-detector joint observations. 
Therefore, we use the aLIGOaLIGOO3LowT1800545 PSD \citep{baltus2021gwtc} (O3 PSD hereafter) as a baseline reference and compute the variations in SNR under both the  aLIGOAPlusDesignSensitivityT1800042 PSD (Aplus PSD hereafter) \citep{Barsotti2018} and the EinsteinTelescopeP1600143 PSD (ET PSD hereafter) \citep{Evans2020} scenarios.

\begin{figure}
    \centering    \includegraphics[width=\columnwidth]{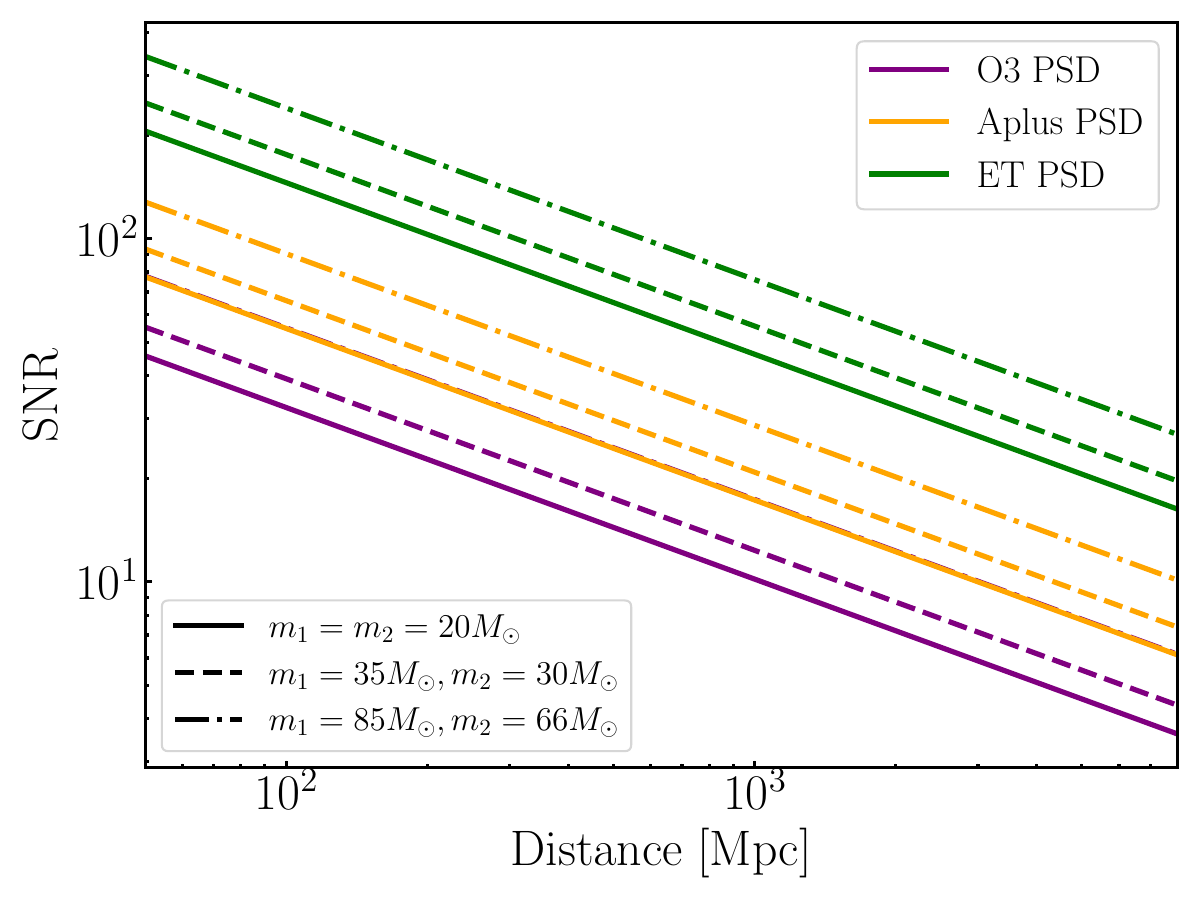}
    \caption{SNR as a function of source distance. 
    The purple, orange and green lines represent O3 PSD, Aplus PSD, and ET PSD, respectively. The solid, dashed and dot-dashed lines correspond to three mass combinations, i.e., $m_1=m_2=20M_{\rm{\odot}}$, $m_1=35M_{\rm{\odot}},m_2=30M_{\rm{\odot}}$, and $m_1=85M_{\rm{\odot}},m_2=66M_{\rm{\odot}}$, respectively.}    \label{fig:SNR_distance_O3_Aplus_ETPSD}
\end{figure}

Based on the selected parameter ranges described above, our calculation results are presented in FIG. \ref{fig:SNR_distance_O3_Aplus_ETPSD}. 
The purple, orange and green lines represent O3 PSD, Aplus PSD, and ET PSD, respectively. The solid, dashed and dot-dashed lines correspond to three mass combinations, i.e., $m_1=m_2=20M_{\rm{\odot}}$, $m_1=35M_{\rm{\odot}},m_2=30M_{\rm{\odot}}$, and $m_1=85M_{\rm{\odot}},m_2=66M_{\rm{\odot}}$, respectively. 
For the three PSD scenarios considered, we have computed the variation of the SNR with luminosity distance for three distinct mass combinations. 
Subsequently, based on the results obtained in this subsection, we will proceed to calculate the distribution of the mismatch.

\subsection{Mismatch Arising from AGN Lensing}
\label{subsec:mismatch}

In this subsection, We first introduce the distribution of the mismatch on the AGN disk for the O3 PSD case. 
Then in the same way, we apply the calculations for the Aplus PSD and ET PSD cases. 

\subsubsection{Mismatch distribution: with O3 PSD}
\label{subsubsec:mismatchdistribution_O3PSD}

A schematic illustration of the lensing system is shown in FIG. \ref{fig:lens_sys} ($a$). 
As indicated by Eq. 
\eqref{eq:mismatch}, the mismatch is determined by parameters $a$, $\phi$, $\iota$, and $M_{\rm{Lz}}$. 
The lensing related parameter ranges adopted in our analysis are as follows: $a \in [500R_{\rm S},5000R_{\rm S}]$, $\phi \in [-25^{\circ},\,25^{\circ}]$ and $\iota \in [65^{\circ},\,115^{\circ}]$. 
This parameter configuration is based on the following considerations: BHs embedded in the disk migrate inward and form binaries at the location of the migration trap \citep{2020ApJ...898...25T}, eventually merging through various mechanisms within this region \citep{2022Natur.603..237S,2025ApJ...990..211S}. 
Reference \citep{2025arXiv250803637V} simulated this process, and their results indicate that the distribution of BBH pair-up locations depends on the AGN disk profile and properties, as well as the mass of the central SMBH \citep{2016ApJ...819L..17B}. 
In the majority of cases, BBHs form at the migration trap, which is typically located at distances ranging from several hundred to several thousand Schwarzschild radii. 
Furthermore, since the distribution of BHs in the accretion disk primarily follows a radial power-law profile \citep{2019ApJ...876..122Y,2022PhRvD.105f3006L}, we treat the inclination angle $\iota$ and the orbital phase $\phi$ as free parameters in our calculation.

\begin{figure*}
    \includegraphics[width=\linewidth]{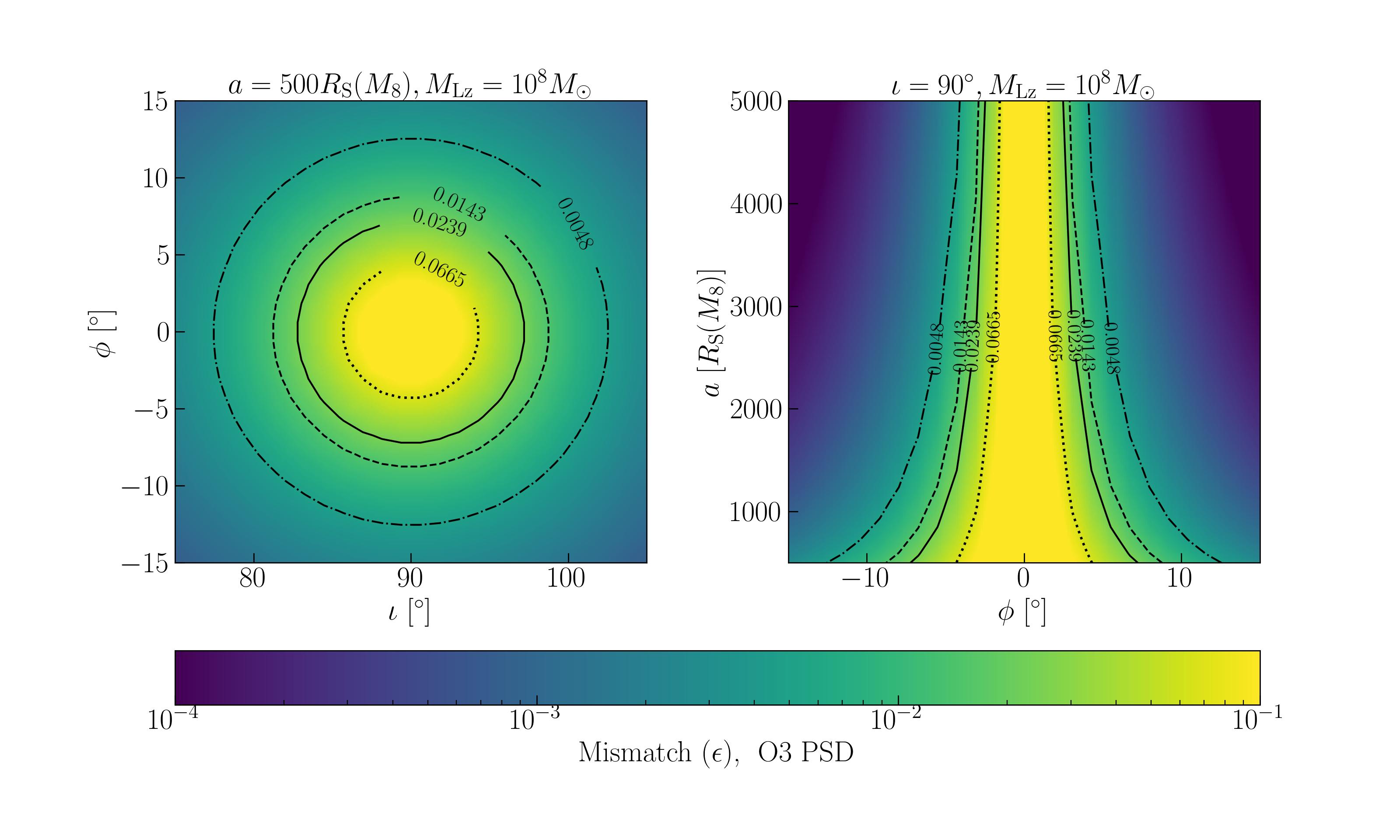}
    \caption{$Left$: Contour plot of the mismatch between lensed and unlensed waveform templates as a function of inclination angle and orbital phase. The solid, dashed, and dash-dotted circle contour line represents mismatch values for $\rm{SNR}$ of $5, 6,$ and $10$, with $M_{\rm{Lz}} = 10^8 M_{\odot}$ and $a = 500R_{\rm{S}}(M_8)$. The dotted circle contour line in this figure indicates the mismatch corresponding to the Einstein criterion. $Right$: Mismatch contours as a function of orbital phase and semi-major axis for a fixed inclination $\iota=90^{\circ}$. The contour levels are the same as in the left panel.}  \label{fig:phi_iota_a500_5000RS_M8_smooth_optimized}
\end{figure*}

We employ the PyCBC python package \citep{2005PhRvD..71f2001A,2012PhRvD..85l2006A,2014PhRvD..90h2004D,2016CQGra..33u5004U,2017ApJ...849..118N,2024zndo..10473621N} to compute the mismatch between the lensed waveform and the corresponding unlensed template waveform \citep{2016APS..APRK14001N,2016CQGra..33u5004U}. 
The unlensed GW waveform template are generated using the IMRPhenomXPHM waveform model \citep{2021PhRvD.103j4056P}. 
In the calculation of the mismatch, except for the change in the source mass, all other parameters are set to their default values.  
The mismatch distribution is shown in FIG. \ref{fig:phi_iota_a500_5000RS_M8_smooth_optimized}. 

In the left panel of FIG. \ref{fig:phi_iota_a500_5000RS_M8_smooth_optimized}, we fix the semi-major axis to $a = 500R_{\rm S}$ and set the lens mass to $10^8M_{\odot}$. 
We investigate how the mismatch varies with inclination angle $\iota$ and orbital phase $\phi$. 
It can be seen that the regions exhibiting significant differences between the lensed and unlensed waveforms are concentrated in a central circular region, corresponding to the well-aligned configurations. 
This significant mismatch is caused by the superposition of two images. 
Given the extremely low probability of such a configuration (for $\phi \approx 0.05^{\circ}$, the probability is $\approx 10^{-7}$), we did not pursue further investigation in this work. 
For readers interested in the case of $\iota \approx 90^{\circ}$ and $\phi \approx 0^{\circ}$, we refer them to Ref. \citep{2025PhRvD.112d3544E} for more details. 
According to Eq. \eqref{eq:threshold}, the SNR of the GW source determines whether the AGN lensing induced waveform deviation can be distinguished from the unlensed one, and thus affects the mismatch threshold. 
Based on the SNR distribution in FIG. \ref{fig:SNR_distance_O3_Aplus_ETPSD}, we mark the contours corresponding to SNR values of 5, 6, and 10 in FIG. \ref{fig:phi_iota_a500_5000RS_M8_smooth_optimized}. 
These SNR values correspond to the O3 PSD case, with $m_1 = m_2 = 20 M_{\odot}$ and source distances of $5000 \rm{Mpc}$, $3000 \rm{Mpc}$, and $1000 \rm{Mpc}$, respectively. 

The dash-dotted circular contour line in FIG. \ref{fig:phi_iota_a500_5000RS_M8_smooth_optimized} corresponds to $\epsilon_{\rm th}$ at an SNR of $\sim 10$. 
This corresponds to a threshold orbital phase of $\phi_{\rm th} \approx 12^\circ$. 
The dashed circular contour line represents $\epsilon_{\rm th}$ at an SNR of $\sim 6$, corresponding to a threshold orbital phase of $\phi_{\rm th} \approx 9^\circ$. 
The solid circular contour line represents $\epsilon_{\rm th}$ at an SNR of $\sim 5$, corresponding to a threshold orbital phase of $\phi_{\rm th} \approx 7^\circ$. 
The dotted circular contour line marks the Einstein radius related mismatch value. 
It is seen that the AGN lensing region determined by the mismatch threshold is generally larger than the Einstein radius. 
The SNR governs the distinguishable parameter space between lensed and unlensed waveforms. 
As the SNR increases from 5 to 10, the parameter space where the lensing effect can be distinguished in principle also expands. 
We will subsequently use the phase angle $\phi_{\rm th}$ corresponding to the mismatch threshold to delineate the region where the AGN lensing effects are distinguishable. 
This region governs the AGN lensing probability. 

In the right panel of FIG. \ref{fig:phi_iota_a500_5000RS_M8_smooth_optimized}, we fix $\iota = 90^\circ$ and set the mass of the SMBH to $10^8M_\odot$ and compute the variation of the AGN lensing induced mismatch as a function of the semi-major axis $a$ and the orbital phase $\phi$. 
The contour lines in the right panel of FIG. \ref{fig:phi_iota_a500_5000RS_M8_smooth_optimized} carry the same meaning as the corresponding line styles in the left panel.
It can be seen that, on the one hand, the region where the lensing effect is significant is concentrated around $\phi = 0^\circ$. As the range of $\phi$ extends outwards from $0^\circ$ towards both sides of the axis, the BBH gradually deviates from the line of sight, resulting in a decrease in mismatch. 
On the other hand, as the semi-major axis increases from $500R_{\rm S}$ to $5000R_{\rm S}$, the $\phi_{\rm th}$ (related to $\epsilon_{\rm th}$) range in which the lensing effect is prominent gradually shrinks. 
This is because when $\iota=90^{\circ}$, this corresponds to an edge-on disk. The angular position of the source is 
\begin{equation}
    \beta = a\sin{\phi}/D_{\rm L},
    \label{eq:betabeta}
\end{equation}
and the dimensionless angular position of the source is 
\begin{equation}
    \eta = \sqrt{\frac{1}{2} r \tan{\phi}\sin{\phi}},
    \label{eq:source_position}
\end{equation} 
in this configuration, where $r=a/R_{\rm S}$. 
It is apparent that a larger semi-major axis imposes stricter geometric constraints on the system, reducing the available parameter space in exchange for the same lensing features (see Eq. \eqref{eq:source_position}). 

If we assume that the AGN disk is circular with an inclination angle of $90^{\circ}$, then, combined with the symmetry of the point-mass lens potential, rotating the AGN disk around the line of sight once would bring all GW sources with orbital phases smaller than $\phi_{\rm{th,max}}$ into configurations where lensing characteristics are distinguishable. 
The parameter $\phi_{\rm{th,max}}$ defines a spherical cap centered on the SMBH at the core of the AGN. 
We have illustrated the geometric configuration of this spherical cap in FIG. \ref{fig:lens_sys} (b). Based on this geometry, we define the AGN lensing probability as the ratio of the solid angle of the spherical cap to the total solid angle $4\pi$. 
The solid angle of the spherical cap is given by Ref. \citep{mazonka2012solid} 
\begin{equation}
    \omega = 2\pi(1-\cos{\phi_{\rm th,max}}).
    \label{eq:spherical_cap}
\end{equation}
The lensing probability therefore takes the form
\begin{equation}
    p({\rm lensing}|\phi_{\rm th,max}) = \frac{1}{2}(1-\cos{\phi_{\rm{th,max}}}).
    \label{eq:probability}
\end{equation}

\subsubsection{Mismatch distribution: with Aplus and ET PSD}
\label{subsubsec:mismatchdistribution_O3PSD}

\begin{figure*}
    \includegraphics[width=\linewidth]{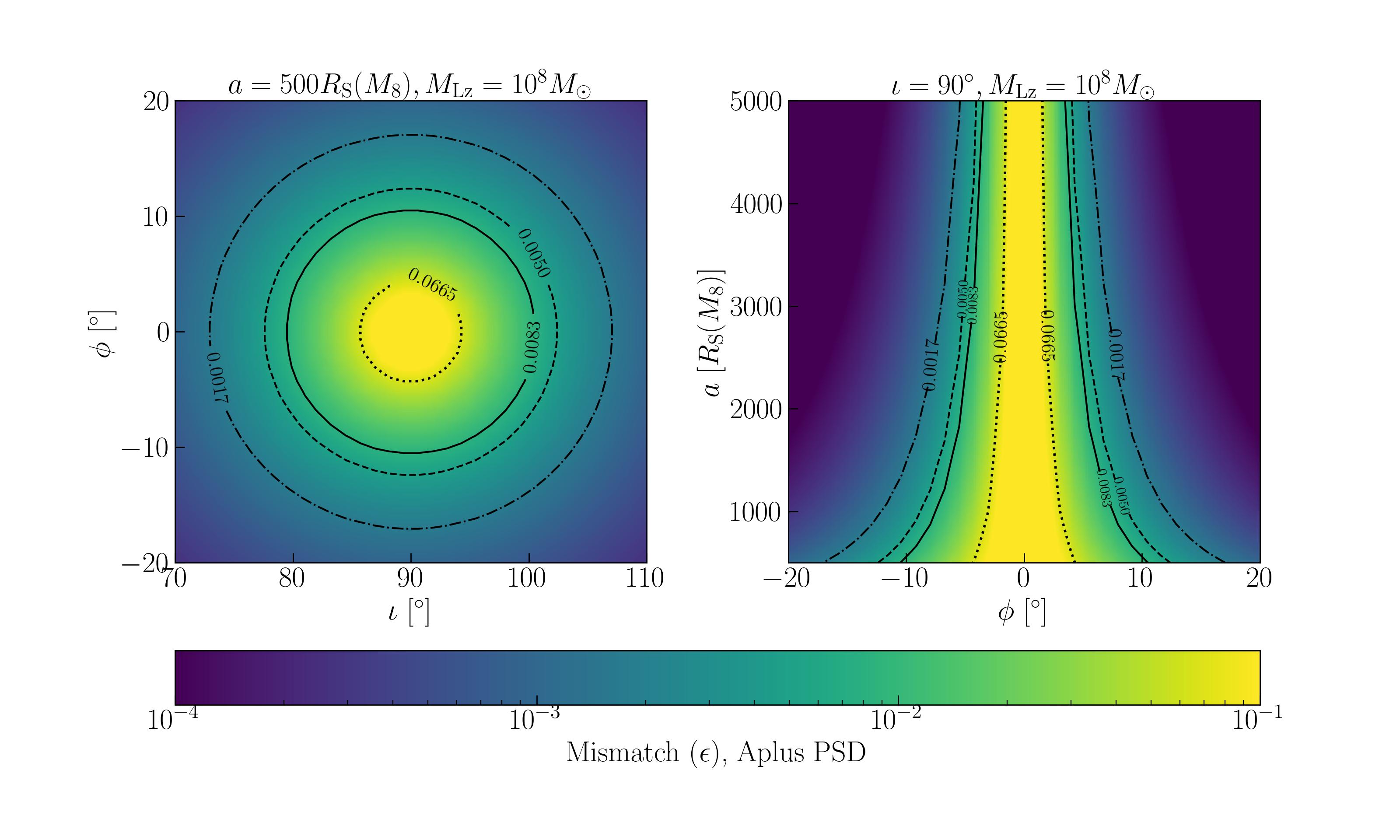}
    \caption{The same as FIG. \ref{fig:phi_iota_a500_5000RS_M8_smooth_optimized} but for Aplus PSD.
    $Left$: The solid, dashed, and dash-dotted circle contour line represents mismatch values for $\rm{SNR}$ of $8, \ 10,$ and $20$, with $M_{\rm{Lz}} = 10^8 M_{\odot}$ and $a_0 = 500R_{\rm{S}}(M_8)$. 
    The dotted circle contour line in this figure indicates the mismatch corresponding to the Einstein criterion. 
    $Right$: Mismatch contours as a function of orbital phase and semi-major axis for a fixed inclination $\iota=90^{\circ}$. The contour levels are the same as in the left panel.
    }  
    \label{fig:Aplus_iotaVSphi_aVSphi_251105}
\end{figure*}

\begin{figure*}
    \includegraphics[width=\linewidth]{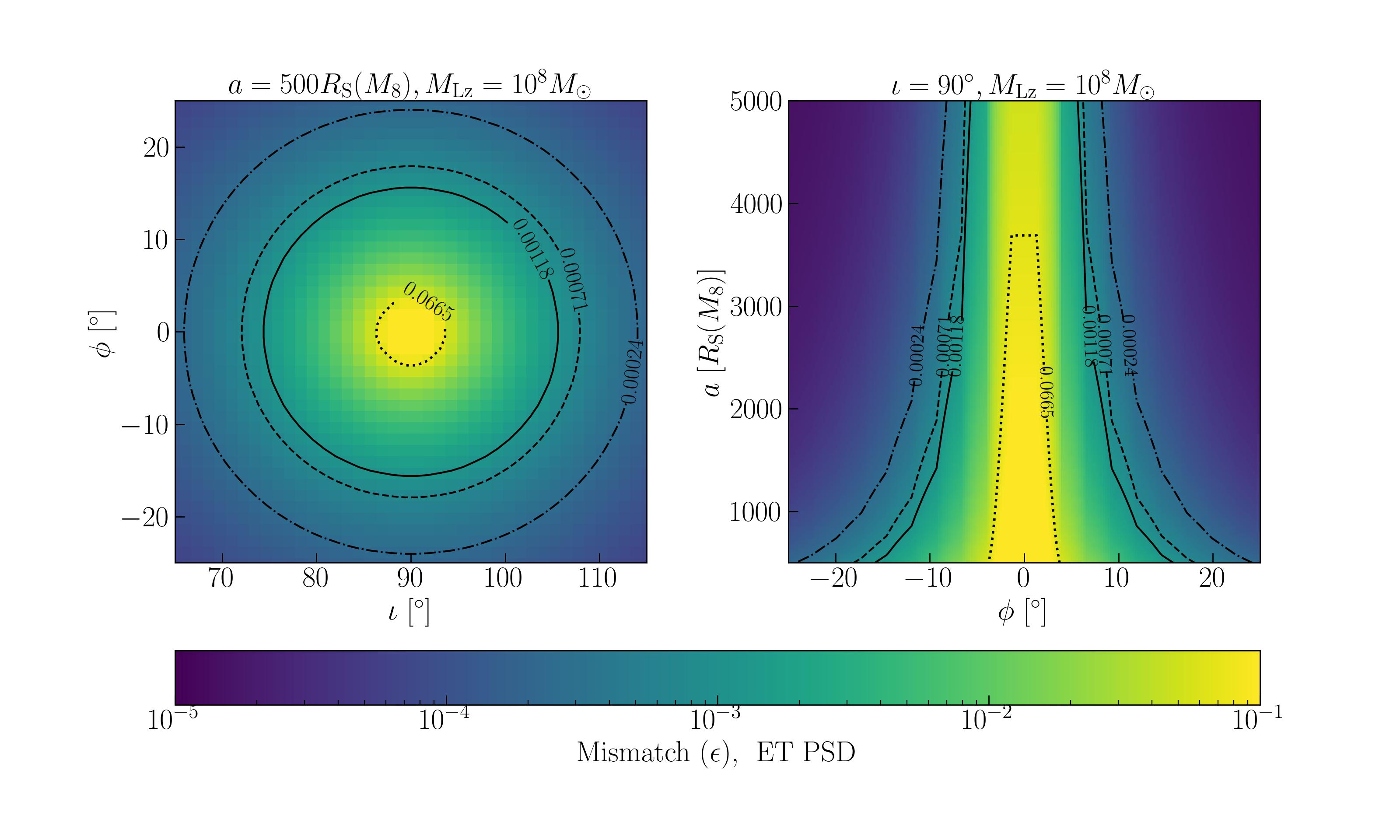}
    \caption{The same as FIG. \ref{fig:phi_iota_a500_5000RS_M8_smooth_optimized} but for ET PSD.
    $Left$: The solid, dashed, and dash-dotted circle contour line represents mismatch values for $\rm{SNR}$ of $20, \ 25,$ and $50$, with $M_{\rm{Lz}} = 10^8 M_{\odot}$ and $a_0 = 500R_{\rm{S}}(M_8)$. The dotted circle contour line in this figure indicates the mismatch corresponding to the Einstein criterion. $Right$: Mismatch contours as a function of orbital phase and semi-major axis for a fixed inclination $\iota=90^{\circ}$. The contour levels are the same as in the left panel.}  \label{fig:ET_iotaVSphi_aVSphi_251105}
\end{figure*}

In this subsection, in order to investigate the influence of different PSDs, we calculate the mismatch distributions for both the Aplus PSD and ET PSD scenarios, based on the results of Section \ref{subsec:Signal-to-Noise Ratio} and following the same procedures and parameter settings as used for the O3 PSD case. 
For the Aplus PSD, the source with the $20M_{\odot}+20M_{\odot}$ 
mass combination yields an SNR of approximately 20 at $1000 \rm{Mpc}$, 10 at $3000 \rm{Mpc}$, and 8 at $5000 \rm{Mpc}$. 
According to Eq. \eqref{eq:threshold}, we calculate the threshold mismatches corresponding to these SNRs, which are represented by the contour lines in FIG. \ref{fig:Aplus_iotaVSphi_aVSphi_251105}. 
The solid, dashed, and dash-dotted circle contour line represents mismatch values for $\rm{SNR}$ of $8, \ 10,$ and $20$, with $M_{\rm{Lz}} = 10^8 M_{\odot}$ and $a = 500R_{\rm{S}}(M_8)$. 
For the ET PSD, the same mass combination yields an SNR of approximately 50 (25, 20) at $1000 \rm{Mpc}$ ($3000 \rm{Mpc}$, $5000 \rm{Mpc}$). 
The corresponding threshold mismatches are indicated by the contour lines in FIG. \ref{fig:ET_iotaVSphi_aVSphi_251105}.

These results demonstrate that at the same luminosity distance, higher detector sensitivity leads to a higher SNR. 
This higher SNRs are associated with smaller mismatch thresholds, which results in a larger distinguishable region for AGN lensing signatures, thereby increasing the AGN lensing probability.
Consequently, we will adopt this mismatch threshold as the criterion for estimating the probability of AGN lensing in the following subsection.

\subsection{Lensing Probability}
\label{subsec:probability}
In this subsection, we first calculate the variation of the threshold orbital phase and lensing probability with respect to the semi-major axis in AGNs, using the Einstein radius as the criterion. 
We then compute the variation of the threshold orbital phase and lensing probability with the semi-major axis under the O3, Aplus, and ET PSD scenarios, using the mismatch threshold as the criterion. 
\subsubsection{Einstein radius criterion}
\label{Einstein_criterion}
In previous work, the Einstein radius is employed as a criterion (as the threshold) to evaluate the significance of lensing effects, serving as a boundary for calculating the probability of lensing events occurring \citep{2025ApJ...979L..27L}. 
In this simple estimation, the Einstein criterion requires that $\beta \leq \theta_{\rm Ein}$. 
From Eqs. \eqref{eq:einstein_angle} and \eqref{eq:betabeta}, this leads to the inequality $r\cos^2 \phi_{\rm Ein} + 2 \cos \phi_{\rm Ein} - r \geq 0$. 
Solving this inequality yields the threshold orbital phase corresponding to the Einstein criterion: $\phi_{\rm Ein} = \arccos{(\sqrt{r^2+1} - 1)/r)}$. 
Substituting this result into Eq. \eqref{eq:probability}, and defining $x = (\sqrt{r^2+1} - 1)/r)$, we obtain $p(\mathrm{lensing}|r) = (1 - x)/2$, which resembles the result in Ref. \citep{2025ApJ...979L..27L}. 
For the purpose of comparison with our results. The orbital phase distribution related to the Einstein criterion as a function of the semi-major axis is shown as a gray dotted line in the upper panel of FIG.  \ref{fig:O3PSD_a_phi_probability_dis1000_dis3000_dis5000_251101}. 
A schematic illustration of the solid angle defined by the Einstein criterion is also shown in FIG. \ref{fig:lens_sys} (b). 
The probability distribution related to the Einstein criterion as a function of the semi-major axis is illustrated by the gray dotted line in the lower panel of FIG. \ref{fig:O3PSD_a_phi_probability_dis1000_dis3000_dis5000_251101}. 

However, the Einstein radius serves only as an approximate boundary for the lensing effect. Further study indicates that the effective lensing region is generally larger than the Einstein radius \citep{2018PhRvD..98h3005L}. 
Reference \cite{2008PhRvD..78l4020L} indicates that the distinguishability of two GW waveforms is determined by the SNR, which in turn depends on the sensitivity of detectors and the intrinsic GW paremeters. 
We focus on the influence of detector sensitivity, the  combination of source mass, and the source distance. 

\subsubsection{Mismatch threshold criterion}
\label{Mismatch threshold criterion: O3 PSD}

In the context of O3 PSD, with the mass combination of $20M_{\odot}+20M_{\odot}$ and other parameters of their default values, the SNR at source distances of $1000\rm{Mpc}$, $3000\rm{Mpc}$, and $5000\rm{Mpc}$ are 10, 6, and 5, respectively. 
Each of these SNRs determines a corresponding mismatch threshold. 
Subsequently, each mismatch threshold corresponds to a threshold orbital phase, which in turn determines the AGN lensing probability at the semi-major axis $a$ for an inclination angle of $90^{\circ}$. 

\begin{figure}
    \centering
    \includegraphics[width=\columnwidth]{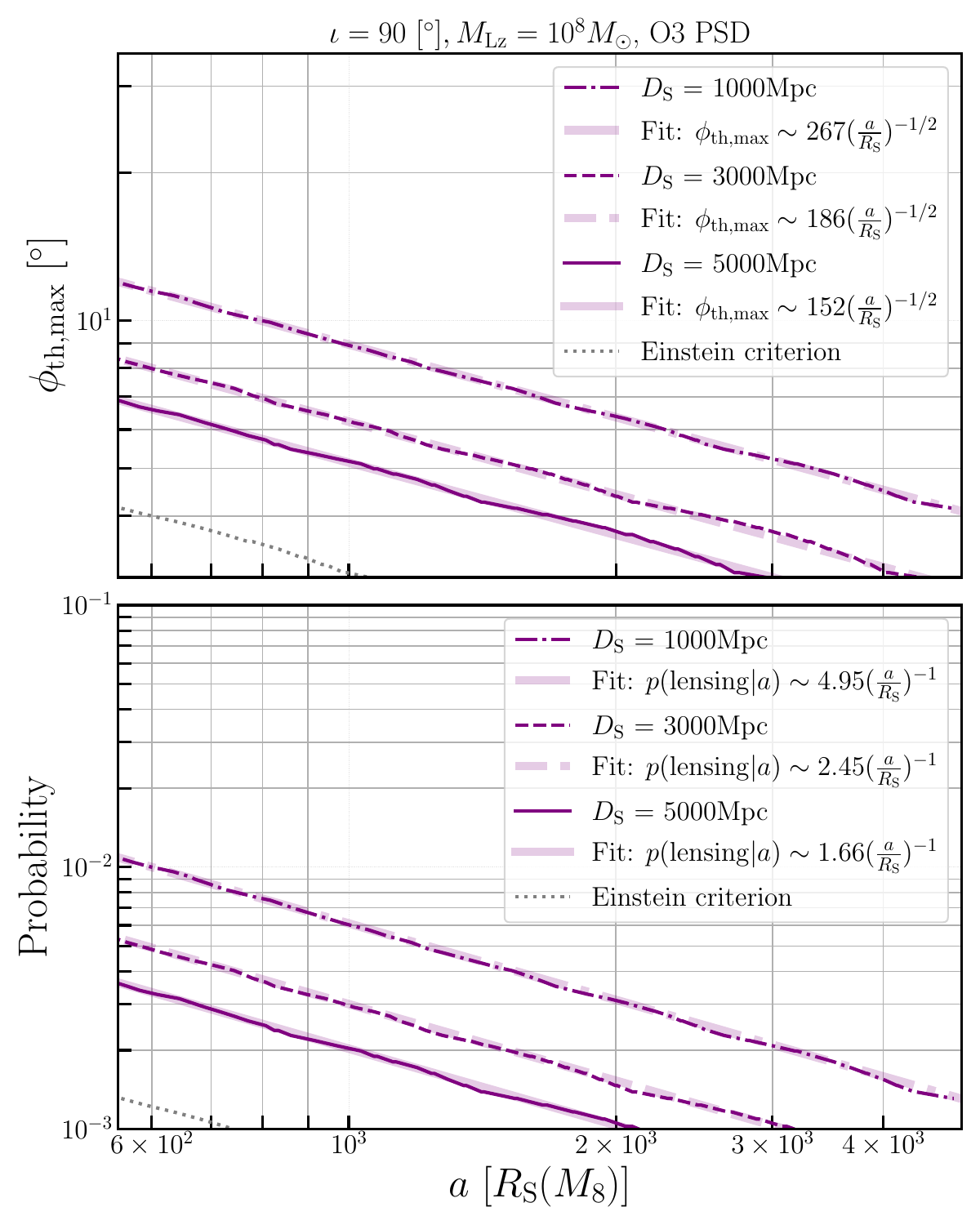}
    \caption{$Upper$: The threshold orbital phase $\phi_{\rm th,max}$ distribution as a function of semi-major axis $a$ with source distanecs of $1000\rm{Mpc}$ (dash-dotted line), $3000\rm{Mpc}$ (dashed line), and $5000\rm{Mpc}$ (solid line). 
    All of the three lines are set $\iota = 90^{\circ}$. 
    The gray dotted line represents the Einstein criterion, which relates to the results of Ref. \cite{2025ApJ...979L..27L}. 
    The shaded line are the power law fitting result of $\phi_{\rm th,max} \sim a^{-1/2}$. 
    $Lower$: Corresponding lensing probability as a function of semi-major axis. 
    The shaded line are the power law fitting result of $p({\rm lensing|a}) \sim a^{-1}$.}   \label{fig:O3PSD_a_phi_probability_dis1000_dis3000_dis5000_251101}
\end{figure}

The relationship between the threshold orbital phase $\phi_{\rm th, max}$ and the semi-major axis $a$ for source distances of $1000\rm{Mpc}$ (SNR$\approx$ 10), $3000\rm{Mpc}$ (SNR$\approx$ 6), and $5000\rm{Mpc}$ (SNR$\approx$ 5) are shown in the upper panel of FIG. \ref{fig:O3PSD_a_phi_probability_dis1000_dis3000_dis5000_251101}. 
The relationship between $\phi_{\rm th,max}$ and $a$ follows the power law of $\phi_{\rm th,max} \sim a^{-1/2}$. 
The shaded regions in the figure correspond to their respective fitting curves, and the SNR determines the specific values of the fitting coefficients. 
As expected, $\phi_{\rm th,max}$ gets smaller when $a$ increases. 
For source distance at $1000\rm{Mpc}$, when the semi-major axis is $a=1000R_{\rm{S}}$, the threshold orbital phase corresponding to the Einstein criterion is approximately 3 degrees, while that corresponding to the mismatch threshold is about 10 degrees; when the semi-major axis is $2000R_{\rm{S}}$, the threshold orbital phase corresponding to the mismatch threshold is approximately 6 degrees. 
In FIG. \ref{fig:lens_sys} (b), we illustrate the cone angles corresponding to the maximum threshold orbital phase derived from the mismatch at $1000R_{\rm S}$ and $2000R_{\rm S}$, as well as the cone angles corresponding to the Einstein criterion. 
At $a=1000R_{\rm{S}}$, the threshold orbital phase is approximately $9^{\circ}$, $6^{\circ}$, and $5^{\circ}$ for distances of $1000 \rm{Mpc}$, $3000 \rm{Mpc}$, and $5000 \rm{Mpc}$, respectively.  
Based on the threshold orbital phase mentioned above. 
It is convenient to employ Eq. \eqref{eq:probability} to evaluate the lensing probability. 
The lensing probability distribution as a function of semi-major axis are shown in the lower panel of FIG. \ref{fig:O3PSD_a_phi_probability_dis1000_dis3000_dis5000_251101}. 

\begin{figure}
    \centering    \includegraphics[width=\columnwidth]{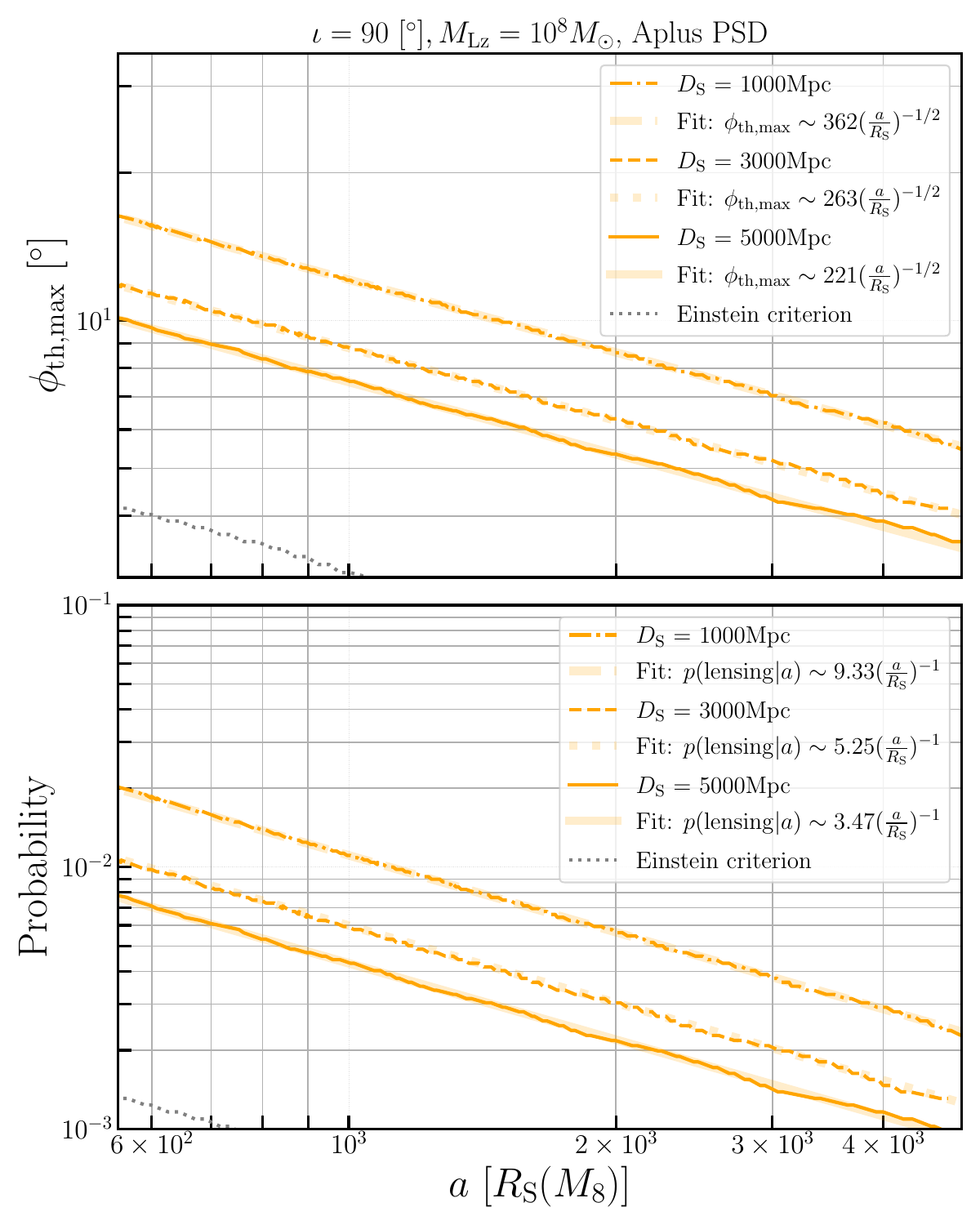}
    \caption{The same as FIG. \ref{fig:O3PSD_a_phi_probability_dis1000_dis3000_dis5000_251101}, but for Aplus PSD. } \label{fig:AplusPSD_a_phi_probability_dis1000_dis3000_dis5000_251101}
\end{figure}

\begin{figure}
    \centering    \includegraphics[width=\columnwidth]{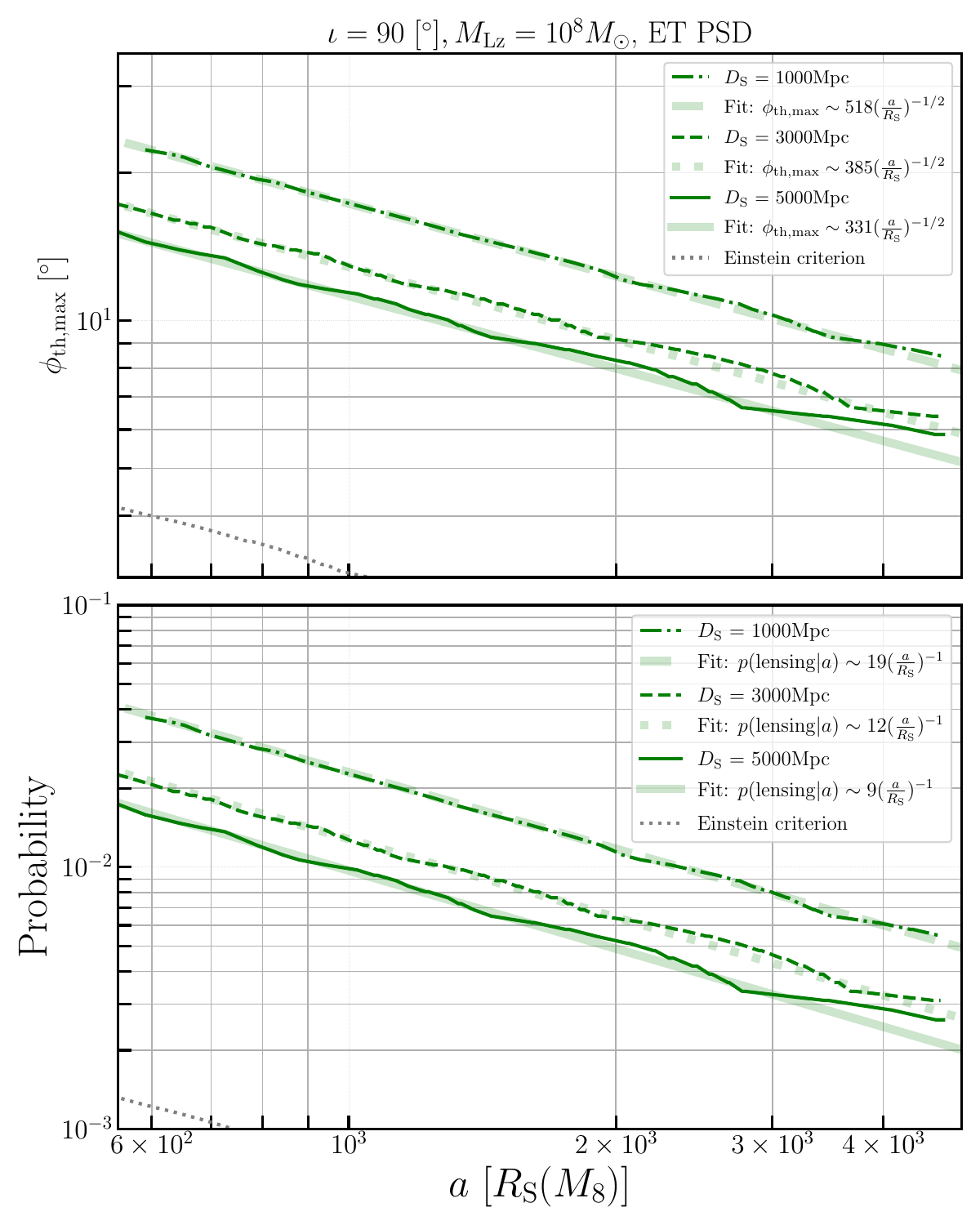}
    \caption{The same as FIG. \ref{fig:O3PSD_a_phi_probability_dis1000_dis3000_dis5000_251101}, but for ET PSD. } \label{fig:ETPSD_a_phi_probability_dis1000_dis3000_dis5000_251101}
\end{figure}

\begin{figure}
    \centering    \includegraphics[width=\columnwidth]{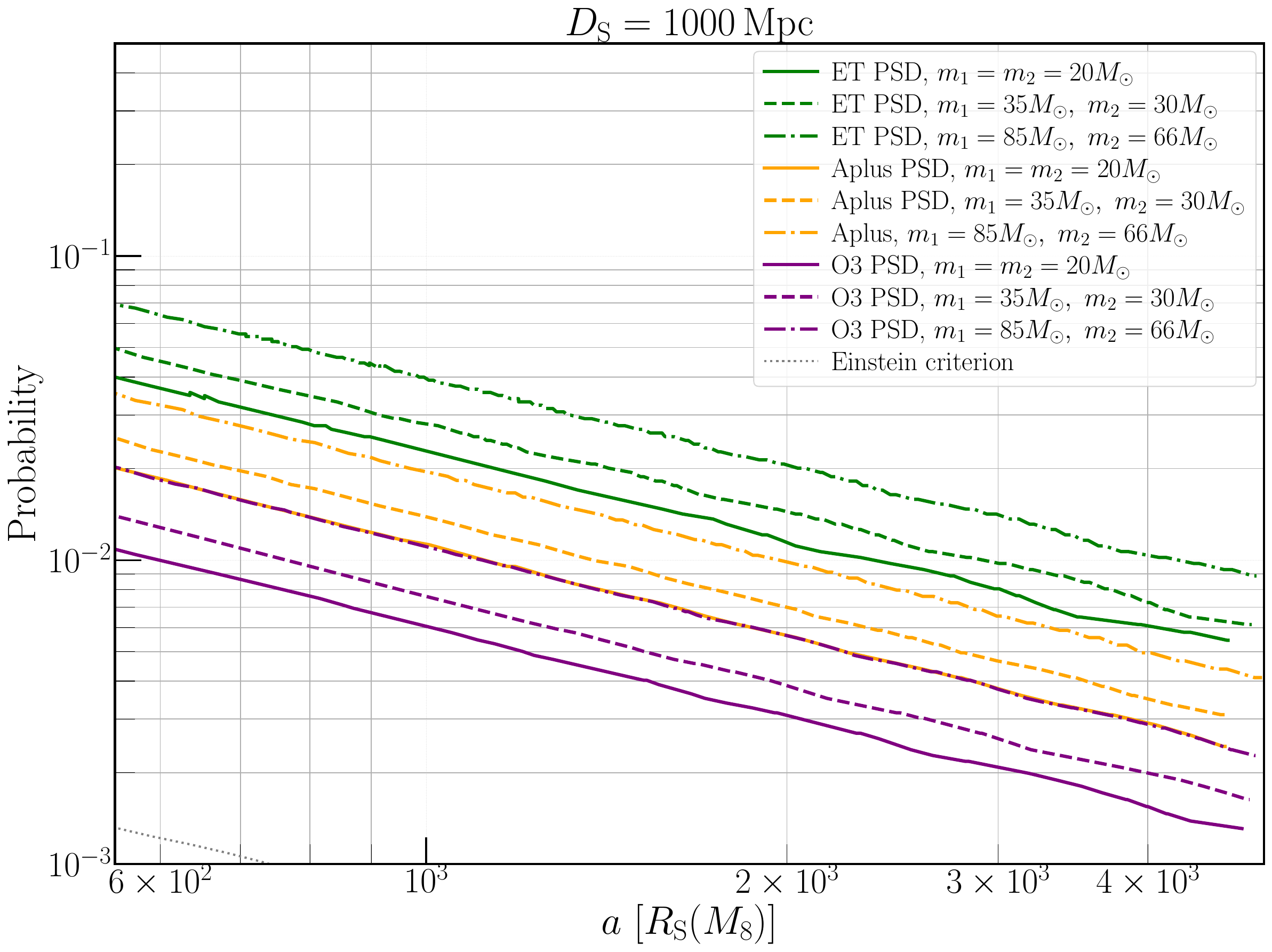}
    \caption{Probability as a function of semi-major axis. 
    Three colors of purple, orange, and green represent O3 PSD, Aplus PSD, and ET PSD respectively, while three line types of solid, dashed, and dash-dotted correspond to three mass combinations: $m_1=m_2=20M_{\rm{\odot}}$; $m_1=35M_{\rm{\odot}},m_2=30M_{\rm{\odot}}$; $m_1=85M_{\rm{\odot}}$, $m_2=66M_{\rm{\odot}}$. The source distance are $D_{\rm S}=1000$ Mpc.} \label{fig:a_phi_probability_increase_dis1000_SNRET46_SNRAplus17_SNRO310}
\end{figure}

Following the same computational procedure as in the O3 PSD case, in FIGs. \ref{fig:AplusPSD_a_phi_probability_dis1000_dis3000_dis5000_251101} and \ref{fig:ETPSD_a_phi_probability_dis1000_dis3000_dis5000_251101} we present the distributions of the threshold orbital phase and their AGN lensing probability as functions of the semi-major axis for the Aplus and ET PSD scenarios, respectively. 
Building on this, we have systematically investigated how different source distances and PSDs lead to variations in SNR, which subsequently determine the threshold orbital phase and ultimately influence the AGN lensing probability. 
Separately, we have also examined how different mass combinations and PSDs result in SNR differences, thereby affecting the threshold orbital phase and finally the AGN lensing probability. 
The corresponding results are presented in FIG. \ref{fig:a_phi_probability_increase_dis1000_SNRET46_SNRAplus17_SNRO310}. 

\section{Conclusion and discussion} 
\label{sec:Conclusion&diskussion}

On the aspect of AGN lensing probability, we propose that the criterion for gravitational lensing should be based on SNR. 
We calculated the lensing probability for BBH mergers in AGN disks through systematic analysis of waveform-template mismatches between lensed and unlensed waveform template banks. 
We summarize our results here. 

We studied the effects of source distances. 
For O3 PSD, under fixed parameters, the AGN lensing probability based on the mismatch threshold is approximately a factor of $\sim 8$, $\sim 4$, and $\sim 3$ higher than that from the Einstein radius criterion at source distances of $1000 \rm{Mpc}$, $3000\rm{Mpc}$, and $5000\rm{Mpc}$, respectively. 
Similarly, for Aplus PSD, the corresponding ratios are $\sim 15$, $\sim 8$, and $\sim 6$. 
Likewise, the corresponding ratios are $\sim 30$, $\sim 16$, and $\sim 13$ for ET PSD. 

We studied the effects of BBH mass combination.
For O3 PSD, the lensing probability based on the mismatch threshold is enhanced by factors of $\sim 10$ and $\sim 15$ for the $35+30 M_{\odot}$ and $85+66 M_{\odot}$ systems, respectively. 
Similarly, for Aplus PSD, the corresponding factors are $\sim 14$, $\sim 18$, and $\sim 26$ for the $20+20 M_{\odot}$, $35+30 M_{\odot}$, and $85+66 M_{\odot}$ combinations. 
Likewise, the corresponding factors for ET PSD are $\sim 80$, $\sim 100$, and $\sim 150$ for the same mass combinations, respectively. 

GW190521 was reported in Ref. \cite{2020PhRvL.125j1102A}. 
Reference \cite{2025ApJ...979L..27L} indicated that, with the lensing probability given by the Einstein criterion, if assuming that all BBHs are GW190521-like and located in the migration trap at $a \simeq 331R_{\rm S}$, one can find that with $\sim 100$ GW events, the probability of observing at least one lensed event is $14\% $. 
In this work, we adopt the mismatch threshold as a criterion to calculate the AGN lensing probability for GW190521-like sources. 
The probabilities are found to be approximately $3\%$ for the O3 PSD, $6\%$ for the Aplus PSD, and $33\%$ for the ET PSD (Note that the source distance is set to $1000\rm{~Mpc}$; for a more realistic distance (e.g., $\sim 3300\rm{~Mpc}$), this percentage would be lower). 
Conversely, the non-detections will place more strict constraints on the fraction of AGN disk BBHs and even the birthplaces of BBH mergers. 
Meanwhile, this work presents the relationship between the lensing probability and the semi-major axis for BBHs in AGN disks. 
This relationship, when combined with the distributions of BBHs within the disk and the AGN population itself, may allow for a better estimation of the AGN lensing event rate. 

We might need to comment that our enhanced rate is ideal. 
In reality, the percentage should be smaller when considering the real detections. 
First, waveform mismodeling \citep{2014PhRvD..89f4037S} in both template generation and detector calibration can obscure subtle lensing signatures. 
Second, data quality issues may potentially affect lensing searches. 
For instance, GW200129 showed some significance in lensing searches \citep{2024ApJ...970..191A}, but in fact might have a significant glitch under subtraction \citep{2022PhRvD.106j4017P}. 
Third, selection bias may also affect the inferred lensing probability. 
Reference \citep{2025PhRvD.111h4019C} shows that, in current matched-filtering searches, lensed gravitational wave signals can lose SNR due to inaccurate parameter recovery and may fail signal-consistency tests, making them less detectable.

As illustrated by the lensing configuration in FIG. \ref{fig:lens_sys} ($a$), when $\iota = 90^{\circ}$, a smaller $\phi$ corresponds to more pronounced lensing features, making it easier to identify the presence of lensing features. 
However, in this case, the lensing probability decreases (see Eq. \eqref{eq:probability}). 
Conversely, as $\phi$ increases, the lensing features become less significant, but the lensing probability increases. 
This poses a significant challenge for the identification of lensed GW events.  
For the case of subthreshold searching \citep{2024SSRv..220...23L,2020arXiv200712709D,2023PhRvD.108j3520E,2021PhRvD.104j3529D,2018arXiv180707062H,2019ApJ...874L...2H,2023PhRvD.107l3014L,2024PhRvD.109b3028G,2023MNRAS.526.3832J,2020PhRvD.102h4031M,2025arXiv250906901C,2025MNRAS.540.2937N,2025ApJ...980..258B,2025MNRAS.542..998L}, where the signal falls below the detection threshold \citep{2024ApJ...970..191A}, the amplitude of the second image is significantly de-magnified. 
This may lead to an inaccurate parameter estimation of the source \citep{2018PhRvD..97b3012N,2018arXiv180205273B}, and the weaker signal from the second image may be sufficiently de-magnified as to fall below the detection threshold (i.e., become subthreshold) and consequently be missed by GW searching pipelines \citep{2021PhRvX..11b1053A,2023PhRvX..13d1039A,2024PhRvD.109b2001A}. 
To justify our revised AGN lensing probability, a dedicated subthreshold targeted search for lensing GW should be performed in the future. 
We stress that this work does not address the identification of lensed gravitational-wave signals. 
For studies on the identification of lensed GW events, we refer the reader to Ref. \cite{2024MNRAS.531..764M}.

GW lensing is also possible for BBHs in a dense star cluster. Therefore, distinguishing the physical origin of BBHs (in a star cluster or AGN disk) through GW lensing alone remains challenging. 
To further constrain the specific formation channel, one needs to consider the other detectable effects of the surrounding environment on the GW sources\citep{2024arXiv240502197G}, 
such as the Doppler effect \citep{2024arXiv241214159S}, eccentricity effects \citep{2016MNRAS.460.3494S,2017ApJ...836...39S}, and strong-gravity effects \citep{2018PhRvD..98f1701Y,2021PhRvD.103d4053A,2024PhRvD.109l4045O,2024MNRAS.535L...1O}.

Regarding the Doppler effect, Ref. \citep{2024arXiv241214159S} pointed out that the relative motion between the source and the lens can cause the observed GW signal from one image to generally appear blue- or redshifted compared to the GW signal from the other image. This velocity-induced differential Doppler shift gives rise to an observable GW phase shift between the GW signals from the different images. 
Regarding eccentric effects, the eccentricity of a binary's orbit depends on the formation history of the binary. 
On the one hand, gravitational radiation causes the orbit of a BBH to circularize continuously \citep{1964PhRv..136.1224P,2021PhRvD.104j4023T}; on the other hand, accretion and dynamical friction can potentially increase the eccentricity \citep{2003ApJ...598..419W,2009MNRAS.395.2127O,2017ApJ...836...39S}. 
Therefore, when we are concerned with identifying the specific formation channel of a BBH, it may be more appropriate to focus on which radial locations exhibit more pronounced eccentric effects. 
Concerning strong-field effects, simulation results from Ref. \citep{2025arXiv250803637V} show that the location of migration traps can be as close as several tens of Schwarzschild radii. 
For BBHs trapped at these locations, strong-field effects are expected to become important \citep{2024PhRvD.109l4045O}. 
Combining these effects with lensing to further investigate the formation channels of BBHs is a particularly interesting endeavor. 
We will explore these scenarios in future work.

\begin{acknowledgments}
We are very grateful to Meng-Ye Wang, Xiang-Li Lei, Hao Wang, Wen-Fan Feng, Xu-Chen Lu, and Bing Zhang for their helpful discussions. This work is supported by the National Key R\&D Program of China (Nos. 2020YFC2201400, SQ2023YFC220007), and the National Natural Science Foundation of China under grant 12473012. W.H.Lei. acknowledges support by the science research grants from the China Manned Space Project with NO.CMS-CSST-2021-B11. The scope of H.L.'s research presented in this article is on the underlying physics and interpretation of the results and was supported by the Laboratory Directed Research and Development program of Los Alamos National Laboratory under project number 20220087DR.

\end{acknowledgments}


\newpage
\bibliography{References} 


\end{document}